\documentclass[twocolumn]{article}

\usepackage[utf8]{inputenc}
\usepackage[margin=1.25in]{geometry}
\usepackage{authblk}
\usepackage{setspace}
\usepackage{graphicx}
\graphicspath{{./figures/}}
\usepackage{subcaption}
\usepackage{amsmath}
\usepackage{amssymb}
\usepackage{booktabs}
\usepackage{hyperref}
\usepackage{float}
\usepackage{xcolor}

\usepackage{bm}

\usepackage[switch]{lineno}       
\usepackage[
  backend=biber,
  style=phys,
  biblabel=brackets,
  sorting=none,
  articletitle=false,
  eprint=true
]{biblatex}
\title{Investigating planar Proca metamaterials in nonlinear (2+1)-Electrodynamics
}

\author[1]{Widervan Morais}
\author[1,2]{S. Strikos}
\author[3]{R. Thibes}
\author[1]{J. A. Helayël-Neto}

\affil[1]{Centro Brasileiro de Pesquisas Físicas}
\affil[2]{Universidade Federal do Rio de Janeiro}
\affil[3]{Universidade Estadual do Sudoeste da Bahia}

\date{2026}

\begin{document}
\twocolumn[
\begin{@twocolumnfalse}

\maketitle

\begin{abstract}
The present work investigates the physics of a planar electromagnetic system incorporating nonlinear effects, along with de Broglie–Proca (dB-P) and Chern–Simons (C-S) mass terms in vacuum, which can be tuned to describe new planar metamaterials (MTMs), potentially suitable for technological applications. As a result, we observe interesting effects generated by a spatially dispersive profile that emerges from non-linearity. This characterizes the three-dimensional QED vacuum as a nonlocal metamaterial. The constitutive relations are explicitly derived.  We show that the dielectric function of the metamaterial (MTM) exhibits spatial dispersion due to the Chern–Simons mass. The dispersion relations are analyzed in terms of applied external electric and magnetic fields as well as in both de Broglie–Proca and Chern–Simons mass parameters. In the special case in which only a background magnetic field is present, this field contributes only to the longitudinal part of the dielectric function and the contribution is driven by the non-linearity coefficient.

\end{abstract}
\vspace{1em}
\end{@twocolumnfalse}
]

\section{Introduction}
Since the 1980s, a large number of new materials have been developed or identified as systems exhibiting strong electron correlations, topological phases or both, giving rise to interesting new physics \cite{Anderson1987, Kivelson:1987efj, Kalmeyer:1987fr,CastroNeto:2007fxn,hasan2010colloquium,Fu:2007moz,teo2008surface,castro2009electronic,liang2016electronic,lai2018weyl,di2023quantum}. As relevant examples in the low-energy spectrum, we can highlight the cases of graphene \cite{CastroNeto:2007fxn, hasan2010colloquium}, topological insulators \cite{Fu:2007moz, teo2008surface}, Dirac and Weyl semimetals \cite{castro2009electronic,liang2016electronic,lai2018weyl} and chiral topological superconductors \cite{di2023quantum} -- all notable for their interesting electronic and structural properties in two and three spatial dimensions, as well as for their potential technological applications. Frequently, the low-energy response of these topological phases can induce modifications to standard electrodynamics \cite{Qi2008Topological}.

On the other hand, nonlocal metamaterials share a deep theoretical connection with topological materials. For example, nonlocality directly modifies the bulk band structure and Berry curvature \cite{chen2025nonlocal}. Concerning their interaction with electromagnetic radiation for example, quite provocative phenomena have emerged such as the fact that electromagnetic radiation emitted by a moving charge particle interacting with non-local material can lead to the Cherenkov effect \cite{gaete2024non}, the quantum Hall effect observed in effective physics of two spatial dimensions \cite{Paschoal:2003xp, kaplan2020fractional}, and several others  which have been reported in recent years \cite{boito2020maxwell}. 
Particular applications in planar physics, with high accessibility to experimental studies and laboratory analysis, aligned with a growing interest in engineering material physics and the development of future technologies, have further driven attention in this field. Therefore, in order to achieve a better understanding on such advanced  material classes and corresponding phenomena, it becomes necessary to develop efficient methods capable of describing their properties at both classical and quantum-mechanical levels, including many-body systems and their possible nontrivial topological aspects. In this context, a quantum field-theoretical approach naturally emerges as a suitable framework to address these challenges \cite{Kleinert:1989kx, Kleinert:1989ky, marino2017quantum, marino2017quantum, Altland_Simons_2023}. 

In this work, we propose a proper investigation of a specific class of bidimensional metamaterials, either natural or carefully synthesized, based on a massive electrodynamics framework constituting a spatially nonlocal medium which we call here de Broglie-Proca metamaterial (dB-P MTM).
The use of lower-dimensional models in quantum field theory has been brought to attention since the works of Jackiw, Templeton and Deser \cite{Jackiw:1980kv, Deser:1982vy}.
Inspired in a recent paper concerning metamaterials and massive electrodynamics in usual three-dimensional space \cite{mikki2021proca}, we consider its restriction to (1+2)-dimensions with a corresponding Chern-Simons (C-S) term and include an additional contribution arising from small perturbations of a non-linear electrodynamics around an external constant background electromagnetic field. In this scenario, the correspondence between Proca fields in vacuum with Maxwell fields in a Proca metamaterial is observed through the impact of the C-S mass and the terms coming from nonlinearity.  As a result, the dielectric function generated by the metamaterial and the resulting spatially dispersive profiles are modified. The recent work \cite{gaete2024non} offers a clue about the effects produced in a three-dimensional space-time scenario with nonlinear terms, which we explore in the current dimensional reduction. In a purely topological case, with a C-S mass term, the wave-vector dependence of the dielectric tensor does not arise, which characterizes the vacuum of the full-fledged model. Hence, additionally to the nonlinear effects, the respective dielectric function depends on two different masses, one associated with the de Broglie-Proca and the other with the C-S term.

We consider the natural covariant dimensional reduction for the mathematical structure and abstract objects present in the standard Maxwell-Heaviside system, particularly driven by the invariant generalized Levi-Civita symbol and corresponding modified duality relations.  While still a component of a corresponding Lorentz second-order tensor, the standard magnetic field becomes a pseudo-scalar with respect to Euclidian space coordinate transformations.  In passing, since the original works \cite{Jackiw:1980kv, Deser:1982vy}, we mention the existence of different approaches to planar electrodynamics in the literature. In references  \cite{Marino:1992xi,  Fernandez:2021dfk,  Gorbar:2001qt, Ozela:2021pse, Marino:2024dmu}, by means of introducing nonlocal operators, the space reduction occurs only with respect to the matter fields content.  Alternatively to nonlinear effects, the use of higher-derivative models \cite{Thibes:2026zal}, enhancing the convergence of Feynman diagrams \cite{Ji:2019phv}, has been discussed in planar electrodynamics in \cite{Lisboa-Santos:2023pwc, Junior:2025xmq}. 

For the reader's convenience,
our work is organized as follows.  In Section {\bf2}, we introduce the proposed model and discuss its general aspects.  In Section {\bf3}, we investigate the system's response towards immersion in an external electromagnetic background, furnishing a metamaterial example.
Particular emphasis on the case of a constant magnetic field configuration is given in Section {\bf 4}. We provide a physical discussion and summarize our main results in Section {\bf 5}.
We use the conventional metric signature $(+1,-1,-1)$ throughout the paper.

\section{General Aspects}
We start off the construction of a field theory in (1+2) space-time dimensions by defining a fundamental density functional $\mathcal{F}$ as a Lorentz and gauge invariant bilinear function of the physical electromagnetic field $(\bm{\mathcal{E}},{\mathcal B})$ given by
\begin{equation}\label{1}
    \mathcal{F}=-\frac{1}{4}G^2_{\mu\nu}=\frac{1}{2}\left(\bm{\mathcal {E}}^2-{\mathcal B}^2\right),
\end{equation}
\noindent with $G^{\mu\nu}$ denoting the usual second rank tensor field strength with corresponding dual $\tilde{G}_\mu=\frac{1}{2}\epsilon^{LC}_{\mu\nu\kappa}G^{\nu\kappa}$. The Lorentz indexes run through $0,1,2$, while $\epsilon^{LC}_{\mu\nu\kappa}$ represents the corresponding totally antisymmetric Levi-Civita symbol. It is important to recall that in (1+2)-dimensions, the magnetic field is a pseudoscalar quantity with respect to Euclidian space transformations.

As shown in \cite{gaete2023vacuum,gaete2023gambini,gaete2024qed, Dib:2025qoy, Spallicci:2024drl, Dib:2026dcu}, if a fluctuating electromagnetic field propagates in the presence of a established (electromagnetic) background, we may split the three-vector potential as the sum of $a^\mu$ and $A^\mu$, where $a^\mu$ stands for the photon field, whereas $A^\mu$ corresponds to the electromagnetic background. This results in the natural decomposition $G^{\mu\nu} = f^{\mu\nu}+F^{\mu\nu}$, where $f^{\mu\nu}$ is the electromagnetic strength tensor associated to $a^\mu$, and $F^{\mu\nu}$ corresponds to the background electromagnetic field strength tensor. The main motivation for this construction in terms of a floating term is to simulate, in particular, the common structure used in laboratories for studying material properties and related phenomena subject to constant background fields. In this context, for the description of a proposed class of planar meta materials, we consider a Taylor expansion of a nonlinear Lagrangian density $\mathcal{L}(\mathcal{F})$ enhanced with Chern-Simon and massive de Broglie-Proca terms, containing the background field information, keeping up to second-order terms in the propagating field, written as
\begin{equation}\label{2}
\begin{aligned}
    \mathcal{L}_{MTM}=-\frac{1}{4}C_1f^{\mu\nu}f_{\mu\nu}+\frac{1}{8}K^{\mu\nu\kappa\lambda}_Bf_{\mu\nu}f_{\kappa\lambda}\\+\frac{1}{2}m\epsilon^{LC}_{\mu\nu\kappa}a^\mu\partial^\nu a^\kappa+\frac{1}{2}M^2a^\mu a_\mu+a_\mu J^\mu,
\end{aligned}
\end{equation}
with the coefficients $C_1$ and $D_1$ fixed in terms of the background electric and magnetic fields, $\textbf{E}$ and $B$, through
\begin{equation}
       C_1=\frac{\partial\mathcal{L}}{\partial\mathcal{F}}\bigg|_{\textbf{E}, B}, \hspace{0,3cm}D_1=\frac{\partial^2\mathcal{L}}{\partial\mathcal{F}^2}\bigg|_{\textbf{E},B}.  
\end{equation}
Still in \eqref{2},
the tensor quantity $K^{\mu\nu\kappa\lambda}_B$ is defined as
\begin{equation}
    K^{\mu\nu\kappa\lambda}_B=D_1F^{\mu\nu}_BF^{\kappa\lambda}_B\,,
\end{equation}
and $J^\mu$ represents a coupling external source.
For generality, we have included all possible terms in \eqref{2} -- the C-S and dB-P masses may be thought as control parameters whose critical limits can be studied later for particular convenient cases.

From the Lagrangian density (\ref{2}), we can derive the corresponding field equations:
\begin{eqnarray}\label{cfe}
    C_1\partial_\mu f^{\mu\nu}-\frac{1}{2}K^{\mu\nu\kappa\lambda}_B\partial_\mu f_{\kappa\lambda}\nonumber\\
    +m\tilde{f}^\nu+M^2a^\nu=J^\nu\,.
\end{eqnarray}
Similarly to $\tilde{G}_\mu$, we have introduced in \eqref{cfe} the dual tensor
$\tilde{f}_\mu=\epsilon^{LC}_{\mu\nu\kappa}f^{\nu\kappa}$.  The reason for attaching the superscript ${LC}$, in notational terms, is to distinguish the Levi-Civita symbol from the electric permittivity, which shall be represented by $\varepsilon$. For a conserved external current, the subsidiary condition $\partial^\nu a_\nu=0$ follows straightforwardly from \eqref{2}.  We also have $\tilde{f}^0=\tilde{f}_0=-b$, $\tilde{f}^i=-\tilde{f}_i=-\epsilon^{LC}_{ij}e_j$, $f_{0i}=e_i$ and $f_{ij}=-\epsilon^{LC}_{ij}b$. In order to find the respective Maxwell equations, it is important to note that the tensor $K^{\mu\nu\kappa\lambda}_B$ can be decomposed into $K^{i00j}_B$ and $K^{i0jk}_B$ in the Gauss equation for $e_i$ and into $K^{0i0j}_B$, $K^{0ijk}_B$ and $K^{jikl}_B$ in the Ampère-Maxwell equation. Writing further $a_\mu=(\phi,a_i)$, this then leads to the modified Maxwell equations
\begin{equation}
    \partial_id_i-mb+M^2\phi=J_0,
\end{equation}
\begin{equation}
    \tilde{\partial}_ih-m\tilde{e}_i+M^2a_i=\partial_td_i + J_i,
\end{equation}
 with
\begin{equation}
    d_i=\varepsilon_{ij}e_j+\zeta_ib ~~~\mbox{and}~ \hspace{0,3cm} h=\mu^{-1}b-\zeta_ie_i,
\end{equation}
representing the displacement and  auxiliary magnetic fields, respectively. Also, we have defined
\begin{equation}
    \varepsilon_{ij}=C_1\delta_{ij}+D_1E_iE_j,
\end{equation}
\begin{equation}
    \mu^{-1}=C_1-D_1B^2,
\end{equation}
\begin{equation}
    \zeta_i=-D_1BE_i,
\end{equation}
with $\varepsilon_{ij}$ denoting the usual electric permittivity, $\mu^{-1}$ the magnetic permeability, and $\zeta_i$ the bi-anisotropic tensor.
Notice that the background electromagnetic fields, denoted by capital letters $(E_i,B)$, follow exactly the same definitions and conventions already stated for $(e_i,b)$, just replacing $f_{\mu\nu}$ by $F_{\mu\nu}$ accordingly. 

By considering a standard expansion in plane wave solutions for the dynamical propagating fields, the complete set of modified Maxwell equations in the absence of sources can  be written in Fourier space as
\begin{equation}\label{3}
    \begin{aligned}
        ik_i\left(C_1e_{0i}+D_1E_iE_je_{0j}-D_1BE_ib_0\right)\\-mb_0+M^2\phi_0=0\,,
    \end{aligned}
\end{equation}
\begin{equation}\label{4}
\begin{aligned}
i\tilde{k}_{i}\left(C_1b_0-D_1B^2b_0+D_1BE_je_{0j}\right)\\+i\omega\bigg(C_1e_{0i}+D_1E_iE_je_{0j}-D_1BE_ib_0\bigg)\\=m\tilde{e}_{0i}-M^2a_{0i}\,,
\end{aligned}
\end{equation}
\begin{equation}\label{5}
    \epsilon^{LC}_{ij}k_ie_{0j}=\omega b_0\,,
\end{equation}
where we have used an extra zero subscript to mark the respective Fourier amplitudes.  Expressing the propagating electric and magnetic fields in terms of the potentials, we write:
\begin{equation}\label{6}
    e_{0i}=-ik_i\phi_{0}+i\omega a_{0i}
\end{equation}
\noindent and
\begin{equation}\label{7}
    b_{0}=i\epsilon^{LC}_{ij}k_ia_{0j}.
\end{equation}

From these equations, we can  find the relations for the scalar and vector potentials. In fact, rewriting (\ref{3}), for $M\neq 0$, we find the scalar potential
\begin{equation}
    \begin{aligned}
        \phi_{0}=\frac{1}{M^2}\bigg(mb_0-iC_1k_ie_{0i}~~~~~~~~~~~~\\-iD_1E_iE_jk_ie_{0j}+iD_1BE_ik_ib_{0}\bigg)\,.
    \end{aligned}
\end{equation}
Further replacing this result into equation (\ref{6}), using (\ref{5}), we obtain the vector potential as:
\begin{equation}\label{10}
    \begin{aligned}
            a_{0i}=\frac{1}{i\omega}\bigg[\delta_{ij}+\frac{C_1}{M^2}k_ik_j\\-\frac{D_1}{M^2}E_jE_kk_ik_k
            +\frac{im}{M^2\omega}\epsilon^{LC}_{jk}k_ik_k\\+\frac{D_1}{M^2\omega}BE_m\epsilon^{LC}_{jk}k_ik_mk_k
            \bigg]e_{0j},
    \end{aligned}
\end{equation}
which can be inverted for $e_{0j}$ as
\begin{equation}\label{e_i}
    \begin{aligned}
           e_{0j}=\frac{1}{i\omega}\bigg[\delta_{ij}+\frac{C_1}{M^2}k_ik_j\\-\frac{D_1}{M^2}E_jE_kk_ik_k
            +\frac{im}{M^2\omega}\epsilon^{LC}_{jk}k_ik_k\\+\frac{D_1}{M^2\omega}BE_m\epsilon^{LC}_{jk}k_ik_mk_k
            \bigg]^{-1}a_{0i}.
    \end{aligned}
\end{equation}

We turn our attention now back to the system characterization by the Lagrangian density (\ref{2}). Note that both the C-S and dB-P mass terms play prominent roles, leading to nontrivial effects on the system, but how exactly this is done is the real question to be answered.  To address this point next, first we express $b_0$ from (\ref{7}) in terms of Faraday-Lenz law (\ref{5}) and then introduce the result into (\ref{10}), allowing the wave equation for electric field in Fourier space to be written as
\begin{equation}\label{21}
    \begin{aligned}
      \bigg[\omega^2\left(C_1\delta_{ij}+D_1E_iE_j+\frac{D_1}{\omega}BE_i\tilde{k}_j\right)\\-\left(k^2\delta_{ij}-k_ik_j\right)\left(C_1-D_1B^2\right)\\-M^2\delta_{ij}+im\omega\epsilon^{LC}_{ij}+D_1BE_j\omega\tilde{k}_i\\-ik_i\bigg(-\frac{m}{\omega}\tilde{k}_j-iC_1k_j-iD_1E_kE_jk_k\\-\frac{i}{\omega}D_1BE_mk_m\tilde{k}_j\bigg)\bigg]e_{0j}=0.  
\end{aligned}
\end{equation}

Note that (\ref{21}) is actually a matrix equation of the form $M_{ij}e_{0j}=0$, whose nontrivial solutions can be found by imposing $\det M=0$. We can simplify the solution finding task by going to the rest-frame, in which case we set $k=0$, to obtain
\begin{equation}
    \begin{aligned}
        \left(C^2_1+C_1D_1\mathbf{E}^2\right)\omega^4 +M^4 \\-\bigg(2C_1M^2+m^2+M^2D_1\mathbf{E}^{2}\bigg)\omega^2=0
        \,.
    \end{aligned}
\end{equation}

From this analysis, we can spot the following key features:
\begin{enumerate}
    \item When both nonlinearity and dB-P terms are removed, such that only the Chern--Simons term remains, the two modes are found to be $\omega^2=0$ and $\omega^2=m^2$. Thus, the nonzero propagating mode is characterized solely by the C-S mass $m$.
    \item When all three contributions, namely nonlinearity, C-S, and dB-P terms, are present, the resulting outcome exhibits an implicit dependence on the magnetic field through the nonlinearity coefficients $C_1$ and $D_1$:
    \begin{equation}
 {\omega^2=\frac{2C_1M^2+m^2+M^2D_1\mathbf{E}^2\pm\sqrt{\Delta}}{2C_1\bigg(C_1+D_1\mathbf{E}^2\bigg)}}
    \end{equation}
with
\begin{equation}
\Delta\equiv \bigg(m^2+M^2D_1\mathbf{E}^{2}\bigg)^2+4C_1M^2m^2
\,.
\end{equation}
    \item In the absence of the external electric field ($\mathbf{E}=0$), we find an effective mass for the photon, containing both C-S and Proca masses, given by
\begin{equation}
\begin{aligned}
\left(m_\gamma^{\mathrm{eff}}\right)^2
&=
\frac{2C_1M^2+m^2}{2C_1^2} \\
&\quad \pm
\frac{m\sqrt{m^2+4C_1M^2}}{2C_1^2}.
\end{aligned}
\end{equation}
\end{enumerate}

Thus, the electromagnetic excitation is characterized by two effective mass scales arising from the interplay between the C-S and dB-P masses, respectively $m$ and $M$. The two mass parameters are coupled through the dispersion relation, giving rise to two distinct massive modes. For fixed
magnetic field, and hence fixed $C_1$, the asymptotic behavior depends on which mass parameter dominates. In the limit $M\gg m$, both effective masses increase approximately linearly with $M$, whereas for $m\gg M$, the two branches exhibit different behaviors: the upper branch increases approximately as $m/C_1$, while the lower branch decreases as $M^2/m$. Furthermore, when either of the two mass parameters vanishes, the remaining mass parameter independently determines the corresponding nonzero massive mode, showing that the two mass-generating mechanisms can act independently.

\section{A metamaterial example in $(1+2)$D}
In this section, we discuss an application of the present system as an important instance in the electromagnetic theory of $(1+2)$D materials.
We start by considering a decomposition of currents and charge densities into external and internal parts \cite{il2013electromagnetic}, 
\begin{equation}
    J(\textbf{k}, \omega)=J_{ex}(\textbf{k},\omega)+J_{ind}(\textbf{k},\omega)
\end{equation}
and
\begin{equation}
    \rho(\textbf{k},\omega)=\rho_{ex}(\textbf{k},\omega)+\rho_{ind}(\textbf{k},\omega)
\end{equation}
respectively. Therefore, assuming a material free of external sources, that is, $\rho_{ex}(\textbf{k},\omega)=0$, $J_{ex}(\textbf{k},\omega)=0$, the current response is given by:
\begin{equation}
    J^{ind}_{i}(\textbf{k},\omega)=\sigma_{ij}\cdot e_j(\textbf{k},\omega)
\end{equation}

\noindent where $\sigma_{ij}$ is the non-local conductivity profile of the material medium.
The metamaterial's field response is given by:
\begin{equation}   
D_i(\textbf{k},\omega)=\varepsilon_{ij}(\textbf{k},\omega)e_j(\textbf{k},\omega)
\end{equation}
\noindent with
\begin{equation}\label{9}
    \varepsilon_{ij}(\textbf{k},\omega)=\delta_{ij}+\frac{i}{\omega}\sigma_{ij}(\textbf{k},\omega)\,,
\end{equation}
representing the equivalent of the normalized dielectric function of the medium.

Next, we establish an equivalence between the Proca and Maxwell theories in our nonlocal MTM, in similar fashion as worked out in \cite{mikki2021proca} for a three-dimensional linear theory with Proca mass. Following \cite{mikki2021proca}, we consider the MMT response
\begin{equation}\label{10.1}
    J_j=-M^2a_j=\sigma_{ij}e_j\,,
\end{equation}
which, by means of \eqref{e_i}, leads to
\begin{equation}
\begin{aligned}
        \sigma_{ij}(\textbf{k},\omega)=-\frac{1}{i\omega}\bigg[M^2\bigg(\delta_{ij}+C_1\frac{k_ik_j}{M^2}\\-\frac{D_1}{M^2}E_jE_kk_ik_k\bigg)+\frac{im}{\omega}\epsilon^{LC}_{jk}k_ik_k\\+\frac{D_1}{\omega}BE_m\epsilon^{LC}_{jk}k_ik_mk_k\bigg]\,,
\end{aligned}
\end{equation}
representing the system conductivity for the current case.  Further replacement of this expression into (\ref{9}) yields
\begin{equation}\label{11}
\begin{aligned}
        \varepsilon_{ij}(\textbf{k},\omega)=\bigg(1-\frac{M^2}{\omega^2}\bigg)\delta_{ij}-\frac{C_1}{\omega^2}k_ik_j\\+\frac{D_1}{\omega^2}E_jE_kk_ik_k+\frac{D_1}{\omega^3}E_kBk_ik_k\epsilon^{LC}_{lj}k_l\\-\frac{im}{\omega^3}k_i\epsilon^{LC}_{lj}k_l\,,~~~~~~~~~~~~~
\end{aligned}
\end{equation}
for the dielectric function of the medium in the current general case containing both $\textbf{E}$ and $B$ as background fields. This result demonstrates the dependence on both background fields arising from the nonlinearity term and their respective coefficients, $C_1$ and $D_1$. In addition, we have a clear extra contribution from the C-S term. In the next section, we proceed with a deeper analysis for the case in which we have only the presence of $B$ as a constant and homogeneous background field. 

\section{Magnetic field background}
A constant magnetic field background is often employed in material
applications and experimental setups. In this section, we consider
only a constant and homogeneous magnetic background $B$ acting on
the system, with the background electric field turned off. In
practical terms, this leads to $K^{i0\kappa\lambda}_B=0$. 
Under these conditions, Eq.~(\ref{11}) reduces to
\begin{eqnarray}
\varepsilon_{ij}(\mathbf{k},\omega)
&=&
\left(1-\frac{M^2}{\omega^2}\right)\delta_{ij}
-\frac{C_1}{\omega^2}k_i k_j
\nonumber\\
&&-\frac{im}{\omega^3}
\epsilon^{LC}_{kj}k_i k_k.
\label{12}
\end{eqnarray}

When the background electric field is turned off, the dielectric
tensor simplifies considerably, retaining explicit dependence on
the nonlinearity coefficient $C_1$ and the Chern--Simons (C-S)
contribution. This simplification is consistent with the relation
between the background fields imposed by Gauss' law, c.f. Eq.~(\ref{3}).

To make the C-S contribution more transparent, we introduce the Euclidian quantities
\begin{equation}
\tilde{k}_i=\epsilon^{LC}_{ij}k_j,
\qquad
\lambda_{ij}=\frac{k_i\tilde{k}_j}{k^2},
\label{lambda_def}
\end{equation}
with $k^2=k_i k_i\neq0$. Using the antisymmetry of the
Levi-Civita tensor, Eq.~(\ref{12}) can be rewritten as
\begin{equation}
\begin{aligned}
\varepsilon_{ij}(\mathbf{k},\omega)
={}&
\left(1-\frac{M^2}{\omega^2}\right)\delta_{ij}
-\frac{C_1}{\omega^2}k_i k_j
\\
&+\frac{imk^2}{\omega^3}\lambda_{ij}.
\end{aligned}
\label{13}
\end{equation}

We introduce further a set of dimensionless operators that facilitate
the decomposition of the dielectric tensor:
\begin{equation}
\omega_{ij}=\frac{k_i k_j}{k^2},
\qquad
\theta_{ij}=\frac{\tilde{k}_i\tilde{k}_j}{k^2}.
\label{projectors}
\end{equation}
In two spatial dimensions, these operators satisfy
\begin{equation}
\omega_{ik}\theta_{kj}
=
\theta_{ik}\omega_{kj}=0,
\end{equation}
\begin{equation}
\omega_{ij}+\theta_{ij}=\delta_{ij}
\,.
\end{equation}
It is clear that $\omega_{ij}$ and $\theta_{ij}$,
respectively, constitute longitudinal and transverse projectors along the $k_i$ space direction, and vice versa for the $\tilde{k}_i =\epsilon^{LC}_{ij} k_j$ direction.

For convenience, we also define the dimensionless symmetric and
antisymmetric operators
\begin{equation}
\begin{aligned}
s_{ij}
&=\frac{k_i\tilde{k}_j+k_j\tilde{k}_i}{k^2},
\\[1mm]
t_{ij}
&=\frac{k_j\tilde{k}_i-k_i\tilde{k}_j}{k^2}.
\end{aligned}
\label{30}
\end{equation}
These definitions imply
\begin{equation}
s_{ij}=\lambda_{ij}+\lambda_{ji},
\qquad
t_{ij}=\lambda_{ji}-\lambda_{ij},
\end{equation}
and, consequently,
\begin{equation}
\lambda_{ij}=\frac{1}{2}(s_{ij}-t_{ij}).
\label{lambda_st}
\end{equation}

Substituting Eq.~(\ref{lambda_st}) into Eq.~(\ref{13}), we obtain
\begin{equation}
\begin{aligned}
\varepsilon_{ij}(\mathbf{k},\omega)
={}&
\left(1-\frac{M^2}{\omega^2}\right)\delta_{ij}
-\frac{C_1}{\omega^2}k_i k_j
\\
&+\frac{imk^2}{2\omega^3}(s_{ij}-t_{ij}).
\end{aligned}
\label{14}
\end{equation}

The operators $\omega$, $\theta$, $\lambda$ form a closed
algebra, as shown in Table~\ref{13.1}.
\begin{table}[htbp]
    \centering
    \begin{tabular}{c|ccc}
        $\cdot$ & $\theta$ & $\omega$ & $\lambda$ \\
        \hline
        $\omega$  & $0$ & $\omega$ & $\lambda$ \\
        $\theta$  & $\theta$ & $0$ & $0$ \\
        $\lambda$ & $\lambda$ & $0$ & $0$
    \end{tabular}
    \caption{Multiplication algebra of the dimensionless operators
    $\omega$, $\theta$, and $\lambda$ in $(1+2)$ dimensions for
    a constant magnetic background.}
    \label{13.1}
\end{table}
While
 $\omega$ and $\theta$ are projectors, $\lambda$ is
nilpotent satisfying $\lambda^2=0$.

The multiplication algebra can be extended to include the
dimensionless symmetric and antisymmetric operators $s$ and $t$.
Using the definitions in Eqs.~(\ref{projectors}) and~(\ref{30}),
we obtain Table~\ref{13.2} below.
\begin{table}[htbp]
    \centering
    \begin{tabular}{c|cccc}
        $\cdot$ & $\theta$ & $\omega$ & $s$ & $t$ \\
        \hline
        $\theta$
        & $\theta$ & $0$ & $-\lambda$ & $\lambda$ \\
        \hline
        $\omega$
        & $0$ & $\omega$ & $\lambda$ & $-\lambda$ \\
        \hline
        $s$
        & $-\lambda^T$ & $\lambda$
        & $\omega+\theta=I$ & $\theta-\omega$ \\
        \hline
        $t$
        & $\lambda^T$ & $\lambda$
        & $\theta-\omega$ & -($\omega+\theta)=-I$
    \end{tabular}
    \caption{Multiplication algebra of the dimensionless operators
    $\omega$, $\theta$, $\lambda$, $s$, and $t$ in $(1+2)$
    dimensions for a constant magnetic background. Here,
    $\lambda^T$ denotes the transpose of $\lambda$.}
    \label{13.2}
\end{table}

The longitudinal and transverse components of the dielectric tensor
can be obtained by projecting Eq.~(\ref{14}) onto the corresponding
projectors. Since
\[
k_i k_j=k^2\omega_{ij},
\]
the longitudinal component is
\begin{equation}
\begin{aligned}
\varepsilon^L
&=\omega_{ij}\varepsilon_{ij}
\\
&=1-\frac{M^2}{\omega^2}
-\frac{C_1k^2}{\omega^2},
\end{aligned}
\label{19}
\end{equation}
whereas the transverse component is
\begin{equation}
\varepsilon^T
=\theta_{ij}\varepsilon_{ij}
=1-\frac{M^2}{\omega^2}.
\label{18}
\end{equation}
The C-S contribution does not modify these diagonal projections,
because $\lambda$ is off-diagonal with respect to the longitudinal
and transverse subspaces. Thus, the nonlinearity affects the
longitudinal response through the term proportional to $C_1k^2$,
while the transverse response retains the Proca contribution.

The dielectric response can therefore be characterized by the
frequency-dependent functions
\begin{equation}
\varepsilon^T(\omega)
=1-\frac{M^2}{\omega^2},
\label{18b}
\end{equation}
\begin{equation}
\varepsilon^L(\mathbf{k},\omega)
=1-\frac{M^2+C_1k^2}{\omega^2},
\label{19b}
\end{equation}
and the C-S coefficient
\begin{equation}
\varepsilon^{\mathrm{C-S}}(\mathbf{k},\omega)
=\frac{imk^2}{\omega^3},
\label{20}
\end{equation}
where the last expression denotes the coefficient multiplying
the dimensionless operator $\lambda_{ij}$.

\begin{figure}[!htbp]
    \centering
    \includegraphics[width=1.0\linewidth]{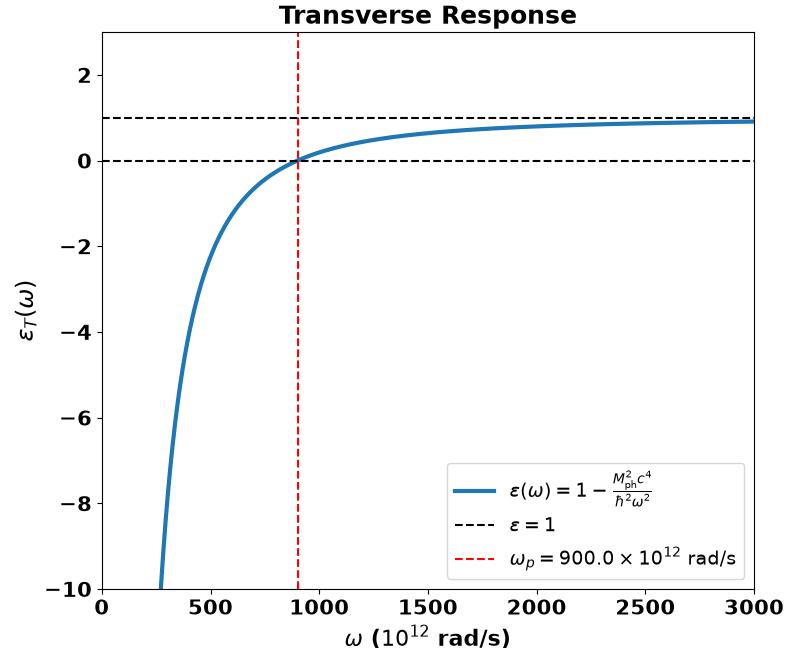}
    \caption{Transverse dielectric function of the nonlocal MTM
    as a function of frequency.}
    \label{fig1}
\end{figure}

\begin{figure}[!htbp]
    \centering
    \includegraphics[width=1.0\linewidth]{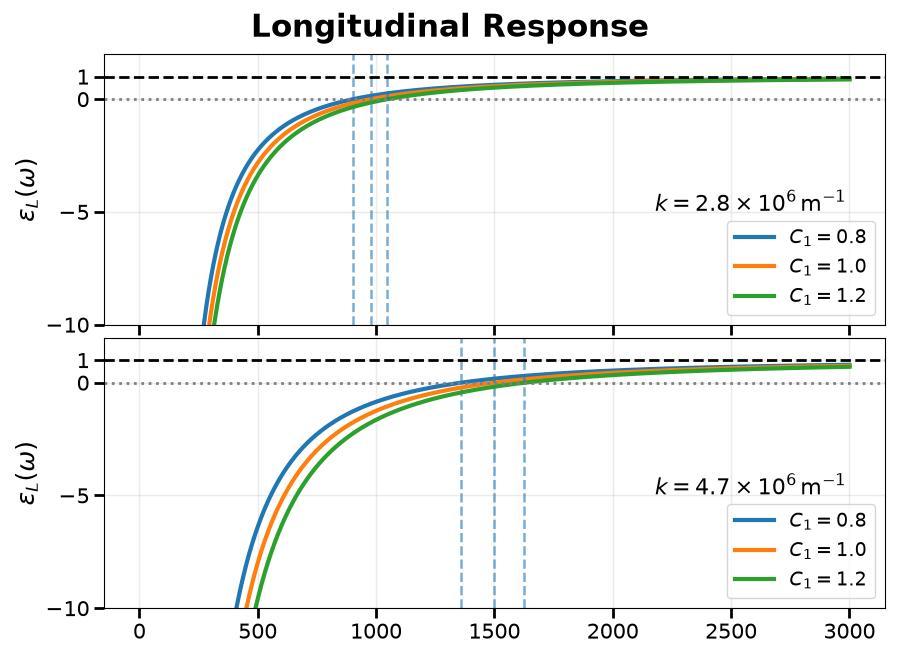}
    \caption{Longitudinal dielectric function of the nonlocal MTM
    as a function of frequency for different values of $C_1$
    and two selected values of $k$.}
    \label{fig2}
\end{figure}

Figures~\ref{fig1} and~\ref{fig2} show the frequency dependence
of the transverse and longitudinal dielectric functions,
respectively, in SI units.
The curves describe the transverse and longitudinal optical
responses as functions of $\omega$ when the Proca mass $M$ is
real. Compared with the results of Ref.~\cite{mikki2021proca},
the nonlinear contribution modifies the longitudinal dielectric
function, while the C-S mass introduces an additional
parity-odd, off-diagonal response.

The Chern--Simons contribution has odd parity due to the presence
of the Levi-Civita tensor in the Lagrangian density,
\begin{equation}
\mathcal{L}_{\mathrm{C-S}}
=
\frac{1}{2}m\epsilon^{LC}_{\mu\nu\kappa}
a^\mu\partial^\nu a^\kappa\,,
\end{equation}
which changes sign under parity transformations in $(2+1)$
dimensions. In the present formulation, this parity-odd character
is manifested in the longitudinal/transverse decomposition of
the dielectric tensor.

Using $\tilde{k}_i=\epsilon^{LC}_{ij}k_j$, the C-S contribution
obtained from Eq.~(\ref{14}) can be written as
\begin{equation}
\varepsilon^{\mathrm{C-S}}_{ij}
=
\frac{im}{\omega^3}k_i\tilde{k}_j
=
\frac{imk^2}{\omega^3}
\hat{e}_{L,i}\hat{e}_{T,j},
\end{equation}
where
\[
\hat{e}_{L,i}=\frac{k_i}{k},
\qquad
\hat{e}_{T,i}=\frac{\tilde{k}_i}{k}
\]
define an orthonormal longitudinal/transverse basis.

Consequently, in the ordered basis $(L,T)$, the C-S contribution
takes the form
\begin{equation}
\varepsilon^{\mathrm{C-S}}
=
\frac{imk^2}{\omega^3}
\begin{pmatrix}
0 & 1\\
0 & 0
\end{pmatrix},
\end{equation}
showing explicitly that the C-S term couples the transverse
component to the longitudinal response. This off-diagonal
contribution is analogous to the transverse response associated
with the Hall effect~\cite{ortix2021nonlinear}, whose conductivity tensor is proportional
to the antisymmetric operator $\epsilon^{LC}_{ij}$. However, since
the present C-S contribution contains only the $L\leftarrow T$
component, it should be regarded as a modified Hall parity-odd
response rather than identified directly with the conventional
Hall conductivity.

Hence,
the complete dielectric tensor can be written as
\begin{equation}
\varepsilon
=
\begin{pmatrix}
\varepsilon_L & \dfrac{imk^2}{\omega^3}\\[2mm]
0 & \varepsilon_T
\end{pmatrix},
\end{equation}
where the diagonal terms contain the parity-even nonlinear
and Proca contributions.

Figure~\ref{fig1} shows the material (T) response computed in the optical range.  It exhibits a Drude-type behavior with resonance frequency characteristics at approximately $0.9\times10^{15}$rad $s^{-1}$~\cite{mikki2021proca}, which corresponds to phonon mass $M=1.054\times10^{-36}$ kg . In Eq.~(\ref{18}), $\omega$ takes values that depend only on the photon mass $M$, and additionally takes values several orders of magnitude smaller than the electron mass energy scale based on the relation
\begin{equation}
\omega_p=m\sqrt{1+\left(\frac{\lambda_{ph}}{\lambda}\right)^2}\,.
\end{equation}

In Table \ref{tab:plasma_frequency}, we have gathered characteristic optical plasma frequencies obtained from the intersections of the longitudinal dielectric response function with the frequency-axis, for different values of the C$_{1}$ coefficient and wave-number.

\begin{table}[!htbp]
    \centering
    \caption{Plasma frequency $\omega_p$ in the optical region, determined numerically from the condition $\epsilon_L(\omega_p)=0$, for different values of $k$ and $C_1$.}
    \label{tab:plasma_frequency}
    \begin{tabular}{ccc}
        \hline
        $k$ ($\mathrm{m}^{-1}$) & $C_1$ &
        $\omega_p$ ($10^{12}\,\mathrm{rad/s}$) \\
        \hline
        $2.80\times10^{6}$ & 0.8 & 902.49 \\
        $2.80\times10^{6}$ & 1.0 & 977.55 \\
        $2.80\times10^{6}$ & 1.2 & 1047.24 \\
        \hline
        $4.71\times10^{6}$ & 0.8 & 1359.14 \\
        $4.71\times10^{6}$ & 1.0 & 1498.86 \\
        $4.71\times10^{6}$ & 1.2 & 1626.62 \\
        \hline
    \end{tabular}
\end{table}

\begin{figure}[!htbp]
    \centering
     \includegraphics[width=1.0\linewidth]{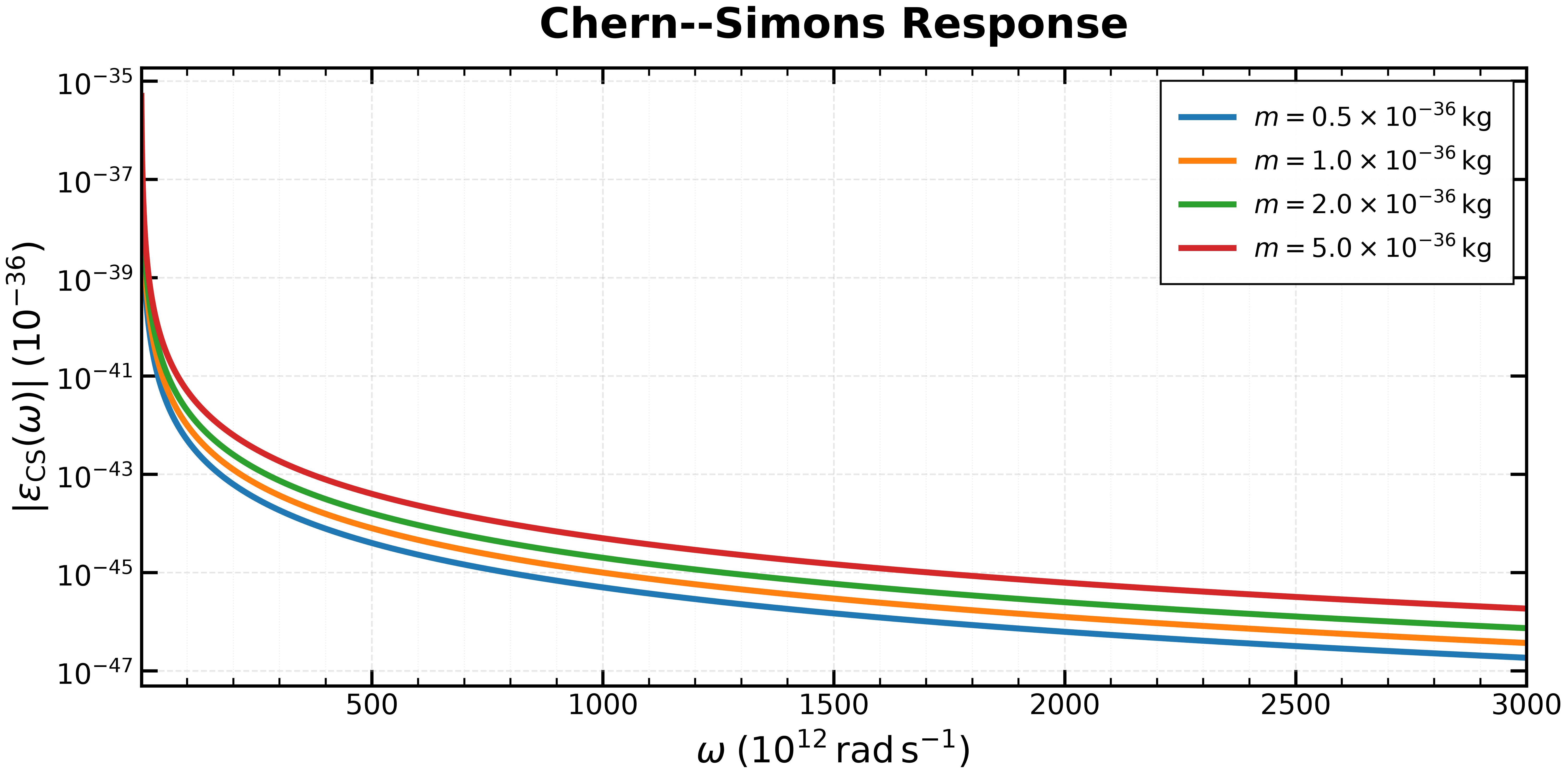}
    \caption{Optical dielectric function modulus of the Chern-Simons (C-S) term of the theory as a function of $\omega$, for two selected values of $k$. Note that a logarithmic-scale was chosen for the vertical axis.}
    \label{fig3}
\end{figure}

The graphic representation of the modulus of relation \eqref{20} is shown in Fig~\ref{fig3},
in which we sample specific mass values for the C-S term behavior on the dielectric function. We observe an asymptotic behavior with a divergence at infinity around the point $\omega=0 \hspace{0.1cm} rads\hspace{0.1cm}s^{-1}$. This is the contribution that the C-S term offers, directly linked to the transverse part. Note that Fig \ref{fig3} shows a very small incremental influence of the C-S term on the dielectric function of the order of $\times 10^{-41}Fm^{-1}$ up to $\times 10^{-36}Fm^{-1}$ when approached from the point of view of planar physics, showing that certain physical properties of this type of materials can be influenced if they are confined in two spatial dimensions.

\section{Discussion and Final Considerations}
We conclude our investigation by discussing further some results of interest. We have observed that the dimensional reduction of the four-dimensional space-time Proca massive model, as proposed in Ref.\cite{mikki2021proca}, aligned with a corresponding C-S term, allowed by dimensional reduction, and nonlinear effects coming from the expansion of a generalized Maxwell electrodynamics around a background field configuration leads to interesting new physics which can be used to describe new planar metamaterials.
For that particular purpose, we have performed a Taylor expansion around the background fields, $\textbf{E}$ and $B$, and we have found that the C-S term and the two background fields all contribute to the constitutive equations with the presence of an extra term. In particular, the addition generated by the background fields occurs through the product of the fields, as seen in Eq.~\eqref{11}.
The new effects become more evident when considering the particular case in which only a background magnetic field $B$ is present, which is the most common case encountered in the laboratory. Here, the expression of the dielectric function (\ref{14}) is shaped by contributions of both the C-S and the non-linearity $C_1$ terms as seen in (\ref{19}) and (\ref{20}). Interestingly, the C-S dielectric function does not represent a projector operator as seen in Table \ref{13.2}. Furthermore, we observe that the contribution of the C-S term to the dielectric function is small from the point of view of effective physics in two spatial dimensions, but not zero. This opens a door to a deeper understanding and future applications of the so-called metamaterials, especially those with planar characteristics, such as Dirac materials, graphene, topological insulators, crystals, and others. Thus, we see that through a rigorous theoretical approach, in the (1+2)D regime, the equivalence between the Proca+C-S+MENL theory and Maxwell's theory in vacuum has been proven to be quite promising regarding the spatially dispersive treatment of a metamaterial.

\subsection*{Acknowledgments}  W.M. acknowledges financial support from Conselho Nacional de Desenvolvimento Científico e Tecnológico (CNPq), Brazil, in the form of a PhD Fellowship. S.S. acknowledges financial support provided by the Fundação Carlos Chagas Filho de Amparo à Pesquisa do Estado do Rio de Janeiro (FAPERJ) through the Programa Pós-Doutorado/Recém-Doutor (PDR – Jovem Pesquisador Fluminense), Process No. E-26/204.725/2024 and the Brazilian Center for Physics Research (CBPF) for providing the research facilities and computational infrastructure used in this work.
R.T. acknowledges CNPq partial funding under grant no 305876/2026-0.
J. A. H.-N. expresses his gratitude to P. Gaete for relevant discussions on the radiation problem in planar Electrodynamics. Thanks are also due to H. Belich for sharing his knowledge on the application of field-theoretical methods to Condensed-Matter systems in lower dimensions.

\printbibliography

\onecolumn

\end{document}